\documentclass{ringb99}
\usepackage{graphics}
\def\etal{{\it et al.\/}}
\def\eg{{\it e.g.}}

\def\cf{{\it cf.}}
\def\ltw{\>\hbox{\lower.25em\hbox{$\buildrel <\over\sim$}}\>}
\def\gtw{\>\hbox{\lower.25em\hbox{$\buildrel >\over\sim$}}\>}
\def \be{\begin{equation}}
\def \ee{\end{equation}}
\begin{document}

\title{Turbulent Particle Acceleration in the Diffuse Cluster Plasma}

\author{Jean Eilek and James Weatherall}
\institute{Physics Department, New Mexico Tech, Socorro NM 87801, USA}

\date{draft:  \today}

\maketitle

\begin{abstract}

{\it In situ} particle acceleration is probably occuring in cluster
radio haloes.  This is suggested by the uniformity and extent of the
haloes, given that spatial diffusion is slow and that radiative losses
limit particle lifetimes.  Stochastic acceleration by plasma
turbulence is the most likely mechanism.  Alfven wave turbulence has
been suggested as the means of acceleration, but it is too slow to
be important in the cluster environment.  We propose, instead, that
acceleration occurs {\it via} strong lower-hybrid wave turbulence.  We
find that particle acceleration will be effective in clusters if only
a small fraction of the cluster energy density is in this form.

\end{abstract}

\section{Introduction}

Diffuse synchrotron emisison from the intercluster plasma (ICM) is
produced by relativistic electrons moving in the cluster magnetic
field.  The origin of these particles is probably easy to explain.
They may have been produced initially by galaxies undergoing active
star formation at early epochs (V\"olk \& Atoyan 1999), or they may be
injected continually  into the ICM by current-epoch galactic winds and
stripping (Petric \& Eilek 1999).  It has also been suggested that
they can be created by active galaxies in the cluster (Feretti \etal\
1995), or as secondaries from a relativistic baryon population
(En\ss lin 1999).  The particles must then diffuse through the
ICM;  diffusion is a slow process (Eilek 1992, Kirk \etal\ 1996), but
given distributed sources the particles should fill a cluster-scale
halo (such as Coma) in the lifetime of the cluster. 

Whatever their origin, however, their maintenance must be
explained.  The radio-loud electrons lose energy by inverse Compton
losses on the microwave background, with a lifetime $\sim 250$ Myr.
There is also some evidence for magnetic fields in excess of $3 \mu$G
in parts of some clusters; such strong fields will shorten the
radiative lifetime for these electrons.  

Thus, it seems very likely that the electrons are undergoing {\it in
situ} reacceleration in the diffuse ICM.  How might this occur?  Shock
acceleration is proposed for many astrophysical situations.  It is not
obvious to us, however, that shocks are common throughout most
clusters.  The ICM is typically a bit warmer than the galaxies, so
only a few galaxies will have localized bow shocks.  Ongoing mergers
of sub-clusters will, indeed, generate perpheral shocks which may lead
to localized particle acceleration (Roettiger 1999, En\ss lin \etal
1998). However, the diffuse radio haloes seen throughout the volume of
some clusters probably require re-acceleration without shocks. 

It follows that stochasatic, turbulent acceleration must be taking
place in radio halo clusters. In particular, it must be fast enough to
offset the radiative losses suffered by the radio-loud electrons. 
Motivated by this question, we are studying turbulent particle
acceleration in the diffuse ICM.  We are particularly interested in
the detailed coupling between the turbulent energy density and the
relativistic particles. It is well-known that  acceleration by
turbulent Alfven waves is slow;  in the cluster  environment it is 
unlikely to be able to offset radiative losses.  We propose that some
of the turbulent energy resides in nonlinear lower hybrid (LH) waves, which
are much more effective at particle acceleration.   In this paper we
summarize our current work on LH wave acceleration.  Details of our
argument are presented in Eilek, Weatherall \& Markovic (1999), and
also in Weatherall \& Eilek (1999).

\section{Turbulent Acceleration, in General}

To begin, we recall the basics of turbulent acceleration competing
with radiative losses.  Turbulent acceleration results in a stochasic
diffusion in momentum space, while radiative losses lead to a direct
flow through momentum space.  The governing equation can be written
(\eg, Borovsky \& Eilek 1986),
\be
{ \partial f \over \partial t} = { 1 \over p^2} { \partial \over
\partial p} \left( p^2 D_p { \partial f \over \partial p} + S p^4 f
\right)
\ee
The first term on the right describes turbulent acceleration:  $D_p$
is the diffusion coefficient (which can be estimated as $( \Delta p)^2
/ \Delta t$, where $\Delta p$ is the mean momentum gain in a collision
of a particle with a wave or wave packet, and $\Delta t$ is the mean
time between collisions. The second term describes radiative losses, 
if the single-particle loss rate is $ d p / dt = S 
p^2$.  Here, $S$ is proportional to the sum of the energy density in
the magnetic field and the microwave background. From this, we can
also estimate an acceleration time for particles at momentum $p$: 
\be
 \tau_{acc}(p) \simeq { p^2 \over D_p(p)}
\ee

Analytic solutions of this equation exist for a steady-state, closed
system with $D_p = D_o p^r$ (Borovsky \& Eilek).  They describe a
distribution function,
\be
f(p) \propto p^2 \exp \left[ - \left( { p \over p_c} \right)^{3-r}
\right] ~,
\ee
which is peaked at the momentum $p_c \simeq ( D_o / S)^{1/(3-r)}$,
where gains balance losses. These solutions are  illustrated in Figure
1 for $r = 3/2$ (which may describe Alfven wave acceleration), and for
$r = 0$ (which may describe lower hybrid wave acceleration). 

\begin{figure}[htb]
\resizebox{3.0in}{2.5in}{\includegraphics{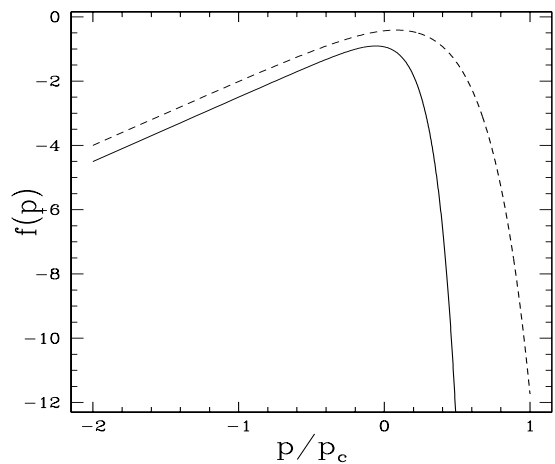}}
\caption[]{Typical particle distribution function for turbulent
acceleration combined with radiative losses.  The dotted line is the
solution for $D_p \propto p^{3/2}$ (a likely solution for Alfven wave
acceleration), while the solid line is for $D_p$ independent of $p$
(the probable solution for LHT acceleration).  From Borovsky \& Eilek
(1986). }
\end{figure}

Our discussion up to here has been set in general terms.  
The heart of the turbulent acceleration problem, however,  lies in
determining the diffusion coefficient, $D_p(p)$.  Computing $D_p$
requires understanding the specific physics of the interaction between
the turbulent plasma waves and the relativistic particles.  In the
rest of this paper we compare two possible situations, Alfven wave
turbulence and lower hybrid wave turbulence.

\section{Alfvenic Turbulence is Slow}

Alfven waves are commonly invoked for wave-particle trapping and
acceleration in the ISM; they are easily generated by macroscopic
(hydrodynamic) processes, and do not damp easily.  They have also been
suggested as an acceleration mechanism in other
diffuse astrophysical plasmas, such as radio galaxies. They cannot,
however, easily explain {\it in situ} acceleration in the ICM.  In
this section we present a brief overview of the physics of Alfven wave
acceleration; Eilek \& Hughes (1991) contains a fuller discussion and
references to the original work. 

Alfven waves are long-wavelength, low-frequency waves which propagate
close to the magnetic field direction.   Most of the  wave energy is in the
transverse 
perturbed magnetic field;  only a small fraction is in the electric
field of the wave (which is capable of doing work on the particles). 

Alfven waves interact with particles through the cyclotron
resonance, when the (Doppler shifted) wave frequency is equal to a
multiple of the particle's gyrofrequency:  $ \omega - k_{\parallel}
v_{\parallel} + s \Omega = 0$ (where $\omega$ is the wave frequency,
$k_{\parallel}$ the component of the wave vector along $\bf B$,
$\Omega$ is the particle's  gyrofrequency  and $s$ an
integer.  Applied to relativistic particles and a typical turbulent
spectrum of Alfven waves, this is a restrictive condition:  a particle
at momentum $p$ sees only one wavenumber, 
$k \simeq m \Omega / p$.  The effect of this is that only a small
fraction of the turbulent Alfven wave energy can be utilized to
accelerate a given particle.   


The wave-particle interaction has been treated by several authors in
the quasi-linear limit.  The result from this work which we need here
is the diffusion coefficient.  This is,
\be
D_A \simeq{ 2 \pi e^2 v_A^2 \over c^2}  { p \over e
B} U_{res}(p) 
\ee
where $U_{res}(p)$ is the turbulent energy {\it at 
wavelengths resonant with particles at $p$}; note in general that
$U_{res} \ll U_A$, the total energy density in Alfven waves.

\section{Lower Hybrid Turbulence is Fast}

Lower hybrid turbulence has not often been proposed in astrophysical
settings, but is well known to exist in the terrestrial ionosphere
where it is associated with particle energization (\eg, Kintner
1992).  In this section
we discuss their effect on relativistic particle acceleration,
summarizing our more detailed argument in Eilek, Weatherall \&
Markovic (1999).  We briefly consider the source of the waves in the
next section (also Weatherall \& Eilek 1999).    

Lower hybrid (LH) waves are short-wavelength, mid-frequency waves which
propagate across $\bf B$.  They have a characteristic frequency,
$\omega_{LH}$, equal to the geometric  mean of the thermal ion and
electron gyrofrequencies.  Most of the wave energy is in the
longitudinal electric field of the wave (and thus is available for
particle acceleration).   When LH 
turbulence becomes strong, it develops intense, localized wave
packets:  field-aligned filamentary regions of high electric field.
Waves inside these wave packets are undergoing ``collapse'' -- rapid
growth of the wave amplitude accompanied with spatial contraction of the
packet.  The collapse preserves the potential drop across the packet
(\eg, Shapiro \etal\ 1993, Melatos \& Robinson 1996).

Particle acceleration occurs through transit time
damping, which happens when a particle crosses the packet in a time
equal to the wave period. We expect that relativistic particles at
all energies will interact with all wave packets, 
once the packet size satisfies ${ l_{\parallel} \omega_{LH} \sim c} $.
Thus, all of the wave packet energy is available for acceleration; and
relativistic particles will be selectively accelerated {\it before}
the nonrelativistic population.  We note that all particles with $v
\simeq c$ will interact with the wave packets once they collapse to
this size;  this is quite different from the Alfven wave resonance
condition discussed above.

\begin{figure}[htb]
 \resizebox{3.0in}{!}{\includegraphics{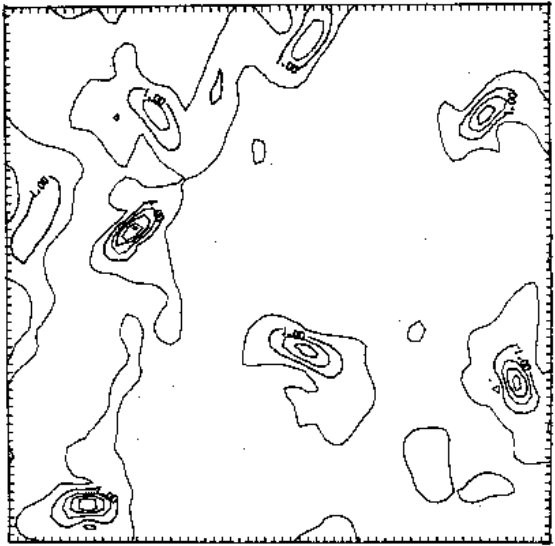}}
\caption[]{Localized wave packets in strong Langmuir turbulence, shown
in contours of the electric field amplitued; from
Weatherall \etal\ (1983).  Strong LH turbulence results in similarly
localized, but more elongated, regions of strong electric field.}
\end{figure}

We are in the process of calculating detailed particle acceleration
efficiencies and diffusion ($D_p$) coefficients in strong LH
turbulence.  As a preliminary step, we have used basic scaling
arguments to estimate what will happen. A particle gains, on
average,  $\Delta p \sim 2 \pi \eta e E_o / \omega_{LH}$ when
transiting the wave packet, if $E_o$ is the packet electric field.
We are currently carrying out numerical simulations to determine the
efficiency $\eta$, as a function of particle energy.  For our present
estimates, we take the $\eta \sim$  1\% based on previous work for
non-relativistic particles (Melatos \& Robinson 1996).  We scale our
estimate to the energy density of the turbulence, $U_{LH}$, use $E_o$
to find the number density of such packets, and from this derive the
typical time it takes a relativistic particle to move between packets.  

From these arguments, we estimate the diffusion coefficient to be 
\be
D_p = D_{LH}\simeq { 8 \pi \eta^2 \over r}   e^2 c \tau_{LH} U_{LH} 
\ee
where $r \ll 1$ is the aspect ratio of the wave packet (transverse
length over parallel length). We note that ${D_p}$ is nearly
independent of 
particle energy (although the factor $\eta$ may be a function of
particle energy).  It is instructive to compare this to the Alfven
wave diffusion coefficient.  Writing the particle energy as $\gamma =
p / m_e c$, we find
\be
{D_{LH} \over D_A } \simeq { 16 \pi \eta^2 \over r} { 1 \over \gamma}
\left( { m_i \over m_e} \right)^{1/2} { c^2 \over v_A^2} { U_{LH}
\over U_{res}(\gamma) }
\ee
This demonstrates that, even in the unlikely case where $U_{res} \sim
U_{LH},$\footnote{Recall that $U_{res}$ is the fraction of the Alfven wave
energy in waves resonant with $\gamma$, where $U_{LH}$ is the total
energy density in lower hybrid waves} inspection of (6) shows that 
$D_{LH} \gg D_A$.  Lower hybrid wave acceleration is thus a much more
effective means of accelerating relativistic particles than is
acceleration by Alfven waves. 

\section{Acceleration in the Cluster Environment}

To discuss particle acceleration in the diffuse cluster plasma, we
pick numbers for a ``typical cluster'':  $B \sim 1 \mu$G; $n \simeq
10^{-3}$cm$^{-3}$; $T \sim 10^8$K (other parameters are given in
Weatherall \& Eilek 1999).  This makes the cluster plasma
weakly  magnetized, with the plasma frequency above the cyclotron
frequency, and the  sound speed above the Alfven speed).  We use these
to describe acceleration by Alfven and by LH waves.

For Alfven wave acceleration, we must choose a wave spectrum.  We
assume a Kraichnan spectrum, $W(k) \propto k^{-3/2}$, with a forward
cascade from the driving scale $\lambda_o \simeq 10$ kpc. 
This wave spectrum gives $U_{res}(p) \propto p^{1/2}$, and the
diffusion coefficient can be written $D_A = D_o p^{3/2}$. We 
scale the total wave energy density, $U_A = \int W(k) dk$ to the total
magnetic field energy density, $U_B = B^2 / 8 \pi$.\footnote{The
overall magnetic field strength is probably the best observational
estimator of the turbulent energy. This scaling allows us to consider
the fraction of that energy going into Alfven or LH waves.} 
When the steady state has been reached, Alfven turbulence supports a
peak particle energy 
\be 
\gamma_c \simeq 9800 \left( {U_A \over U_B} \right)^{2/3}
\ee
and takes 
\be
\tau_{acc} \simeq 2.7 \times 10^8 \left( {U_B \over U_A} \right)^{2/3}
\quad {\rm yrs}
\ee
to reach this energy.   Thus, if $U_A \simeq U_B$, we see that
Alfvenic turbulence can just manage to account for the synchrotron
emission we see.  (For comparison, a particle at $\gamma = 10^4$
radiates at 425 MHz in a $\mu$G field.)

For lower hybrid wave acceleration, we do not need to specify the
turbulent wave spectrum, but only the energy density in waves.  We
find the steady spectrum peaks at
\be
\gamma_c \simeq 1.6 \times 10^7 \left( {U_{LH} \over U_B}
\right)^{1/3}
\ee
which is well above the radio range if $U_{LH} \simeq U_B$.  This peak
is reached in 
\be
\tau_{acc}(\gamma_c) \simeq 1.7 \times 10^7 \left( {U_B \over U_{LH}}
\right)^{2/3} \quad {\rm yrs}
\ee
while the acceleration time for radio-loud electrons (with $\gamma_r
\sim 10^4$, say) can be considerably shorter:
\be
\tau_{acc}(\gamma_r) \simeq 2.1 \times 10^6 { U_B \over U_{LH}} \quad
{\rm sec}
\ee
Thus, even a very low level of LH turbulent energy, $U_{LH} \ll U_B$,
can still easily account for the radio-loud electrons in the cluster.

Our numerical comparison here recovers the result of equation (6):
lower hybrid waves are much more effective at particle acceleration,
for a given level of wave  energy, than are Alfven waves.

\begin{figure}[htb]\hspace{-0.7in}
\resizebox{4.2in}{!}
  {\includegraphics{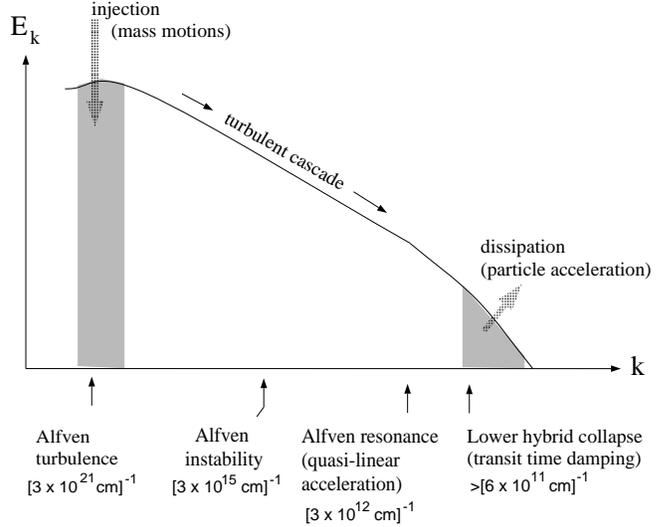}}
\vspace{-2.25in}
\caption[]{An illustration of the possible energy flow from
large-scale hydrodynamic drivers, through Alfven waves undergoing a
forward cascade, to wave dissipation.  Alfven waves will undergo
modest dissipation when they reach wavenumbers resonant with the
relativistic particles, and they will also undergo dissipation through
nonlinear generation of LH waves.  In this cartoon we have suggested
that LH generation ends the cascade;  however that remains to be
verified with numerical simulations presently underway.}
\end{figure}

\section{Connections to Radio Haloes?}
 
To summarize:   we propose that particle acceleration in the ICM
proceeds {\it via} strong lower hybrid wave turbulence.   We have
shown that LHT acceleration is much faster than Alfven wave
acceleration, given similar levels of turbulent energy density in the
two wave forms.   

How are LH waves generated?  They are easily generated by streaming
instabilities (\eg, Galeev, Malkov \& V\"olk, 1995).  As is often the
case, the particle beams necessary for streaming instabilities are not
obviously generated in diffuse 
astrophysical plasmas such as the ICM, and one must look to
hydrodynamic generation.  This is also possible, as LH waves can be
generated by nonlinear mode coupling between lower frequency MHD
waves.   Because Alfven waves are easily generated and not strongly damped
in the cluster setting, we are investigating them as a source of LH
turbulence.  In Weatherall \& Eilek (1999) we present preliminary
numerical work showing that large amplitude Alfven waves do, indeed,
generate lower hybrid waves (through parametric excitation), and that
the LH waves do undergo collapse to produce localized, intense wave
packets.   We thus suspect that normal hydrodynamic processes in the
cluster will generate some level of LH waves. (Figure 3 shows a
cartoon of energy flow in a possible turbulent cascade).  Our argument
above shows that only a modest level of LH turbulence can be very
effective at accelerating relativistic particles, so that we can
afford inefficiency in the coupling between the large-scale turbulent
drivers and the small-scale LH wave turbulence.

\begin{figure}[htb]
\resizebox{\hsize}{3.0in}{\includegraphics{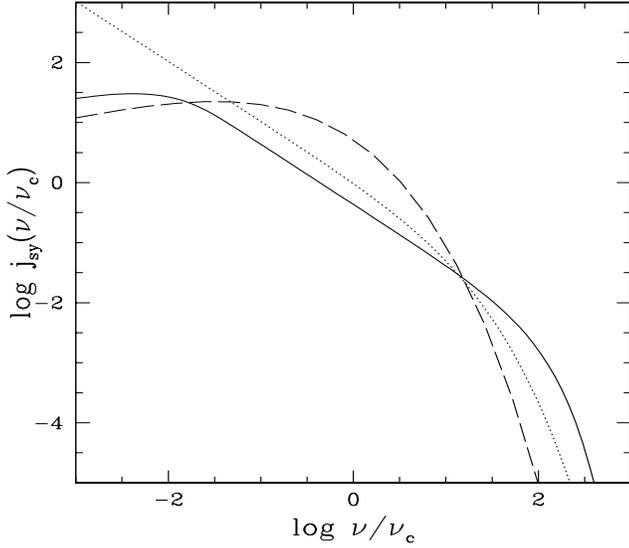}}
\caption[]{Examples of broad synchrotron spectra from narrow particle
distributions.  The fraction of the source volume filled with magnetic
field at strength $B$ is denoted by $g(B)$.  The solid line has $f(p)
\propto \delta(p - p_o)$, and  $g(B) \propto B^{-3}$ for $B_1 < B <
B_2$.  The dotted line has $f(p) \propto p^2  
e^{-p/p_o}$, and $g(B) \propto B^{-3}$ for $B < B_2$ ($B_1$, $B_2$ are
arbitrarily chosen cutoffs).  The dashed line has $f(p)
\propto \delta(p - p_o)$, and $g(B) \propto B^{1/2} e^{-B / B_o}$.
Taken from Eilek \& Arendt (1996). 
}
\end{figure}

What about the distribution function (DF) of particle energies?  We
noted above  that any 
turbulent acceleration mechanism, if let run to steady state in a
closed system with radiative losses, will lead to a particle DF 
that is peaked rather than power law.  At first blush this seems to
violate observations which suggest a broader, probably power-law DF.
This is, however, not a problem:  the magnetic field is almost
certainly 
inhomogeneous ({\it c.f.} Eilek 1999 for arguments on field
filamentation).  Synchrotron emission from a peaked peaked particle DF
in a nonuniform field will lead to a broad photon spectrum.  Eilek \&
Arendt (1996) calculated this for a choice of particle and field
distributions.  Figure 4 shows some examples of broad synchrotron
spectra produced by a narrow particle DF.  In addition, the
particle DF itself may be field dependent. The 
critical energy $p_c$ will reflect the net field seen by the particle
in its lifetime, so that an inhomogeneous magnetic field will lead to
a range of $p_c$ values.  This will also lead to a broadening of the final
photon spectrum.  

Finally, we might ask, how common are radio haloes in clusters?  We
suspect that all clusters contain magnetized plasma, and also have
relativistic particles injected from galaxies in some way or another.
The simple conclusion is that the ICM in all clusters is a source of
diffuse synchrotron emission.  However, it is becoming clear that both
particle acceleration and magnetic field strength (\cf\ Eilek 1999)
are sensitive to the level, and nature, of turbulence in the ICM.
The bolometric  synchrotron emissivity is, of course,
\be
j_{sy} \simeq { 4 \over 3} n_e c \sigma_T \langle \gamma \rangle^2 B^2
\propto u_e^2 B^2 
\ee
where $\langle \gamma \rangle$ is the mean energy of the
relavistic electrons, $n_e$ is their number density, and $u_e = n_e
\langle \gamma \rangle$ is their energy density.  Each of these
quantities, as well as the magnetic field, is determined in part by
the underlying turbulence.  Thus, 
synchrotron emission is merely an indirect tracer of the turbulence, and
one with a very sensitive gain factor.   Perhaps both the turbulence
and its gain factor varies from cluster to cluster, and we have only
detected the brightest haloes and relics up to now.

\begin{acknowledgements}  We are grateful to Tomislav Markovic and
Andreea Petric for ongoing converations about radio haloes and lower
hybrid wave physics. 
This work was partially supported by NSF grant AST-9720263.
\end{acknowledgements}

{}

\end{document}